\documentclass[prb,twocolumn]{revtex4}
\usepackage{graphicx}
\usepackage[ansinew]{inputenc} \usepackage[english]{babel}
\usepackage{pst-all}
\DeclareGraphicsExtensions{.pdf,.ps,.eps}

\newcommand{\Bf}[1]{\mathbf{#1}}
\newcommand{\Rm}[1]{\mathrm{#1}}

\newcommand{\grafico}[4]{\begin{figure}[ht!] \centering
			\includegraphics[scale=#2,clip,keepaspectratio]{#1}
			\caption{\footnotesize{#3}} \label{#4}
			\end{figure} }

\newcommand{\eq}[2]{\begin{equation}#1\label{#2}\end{equation}}
\newcommand{\eqarray}[2]{\begin{eqnarray}#1\label{#2}\end{eqnarray}}
\newcommand{\la}{\langle} \newcommand{\ra}{\rangle}

\begin{document}

\title{Conductance through an array of quantum dots}\label{titulo}
\author{A. M. Lobos}\email{lobosa@cab.cnea.gov.ar}
\affiliation{Centro At\'omico Bariloche and Instituto Balseiro,
Comisi\'on Nacional de Energ\'{\i}a At\'omica, 8400 Bariloche, Argentina}
\author{A. A. Aligia}
\affiliation{Centro At\'omico Bariloche and Instituto Balseiro,
Comisi\'on Nacional de Energ\'{\i}a At\'omica, 8400 Bariloche, Argentina}
\date{\today}

\begin{abstract}
We propose a simple approach to study the conductance through an array of
$N$ interacting quantum dots, weakly coupled
to metallic leads. Using a
mapping to an effective site which describes the low-lying excitations
and a slave-boson representation in the
saddle-point approximation, we calculated the conductance through the
system. Explicit results are presented for $N=1$ and $N=3$: a linear array and an isosceles triangle.
For $N=1$ in the Kondo limit, the results are in very good agreement
with previous results obtained with numerical renormalization group (NRG).
In the case of the linear trimer for odd $N$, when the parameters are such that electron-hole symmetry is
induced, we obtain perfect conductance $G_0=2e^2/h$.
The validity of the approach is discussed in detail.


\end{abstract}
\maketitle

\section{Introduction}

Potential technological applications in electronic devices has led to intense
research in electronic transport in nanoscale systems. In particular, a quantum dot (QD)
acts as a single electron transistor and in addition, the many-body physics
of the ``single impurity" Kondo
effect has been clearly observed in transport experiments.\cite{gold,gold2,cron} More
recently a system of two QD's has been studied, which provides an experimental
realization of two Kondo ``impurities" interacting with a metallic
host.\cite{craig} Also,
linear arrays of 15 QD's have been fabricated to investigate its electronic
properties,\cite{kou} and the conductance through a linear trimer of QD's has
been investigated in the Coulomb blockade regime.\cite{wau}
Systems of coupled QD's are also of interest because of their
possible application in quantum computation.\cite{loss}

Theoretically, the conductance through one QD at equilibrium has been clarified by
different studies of the impurity Anderson model, in particular those using the
highly accurate Wilson's numerical renormalization group (NRG).\cite{nrg1,nrg2,campo}
For moderate values of the Coulomb repulsion $U$,
perturbation theory in $U$ (PTU) provided good
results.\cite{side,pc,lady,oguri,oguricm}
In particular, an approach that interpolates between the second-order
results for the self energy and the atomic limit,\cite{side,pc,lady}
provided accurate results, as has been shown by direct comparison
with exact diagonalization in small rings.\cite{pc}
Renormalized \cite{he2} and interpolative \cite{none} PTU have also been used in the
non-equilibrium case.
At zero temperature, the method of exact diagonalization plus embedding (EDE)
was successfully used.\cite{torio,torio2,wily} This method is based on solving exactly
some part of the system, which includes all interactions, and embedding it in
the rest of the system. A detailed study of the approximation and its range of
validity was given by Chiappe and Verges.\cite{chi} A brief description of the
PTU, EDE and slave boson approximation is contained in Ref. \onlinecite{rev},
where the approaches are applied to describe scanning tunneling spectroscopy of
different systems with magnetic impurities on Cu or noble metal (111) surfaces.

Transport through a system of two QD's, one of them coupled to two
conducting leads has been studied using NRG.\cite{corna,zit} More recently NRG results
for three \cite{ogu3d} and four \cite{nisi} QD's on a line have been reported.
These two works deduce the phase shifts for even and odd parities from the effective
non-interacting Hamiltonian that describes the strong coupling fixed point. For
Fermi liquids with inversion symmetry, the conductance can be expressed in terms
of these two phase shifts.\cite{kawa}
In spite of these studies,
for several QD's or systems with low symmetry,
application of the NRG becomes impractical and one has to
resort to other approximate techniques. The EDE has been applied recently to several
problems involving more than one QD.\cite{busser,mar1,mar2}. In particular,
B{\"u}sser {\it et al.} studied the conductance through a linear array of
QD's.\cite{busser} Remarkably, they find that for an odd number $N$ of QD's except
$N=1$, the conductance $G$ vanishes in the electron-hole symmetric case (EHSC). For $N=1$,
$G$ as a function of gate voltage $V_g$ reaches the maximum at the EHSC with the ideal value
$G_0=2e^2/h$ in agreement with previous
studies.\cite{nrg1,nrg2,campo,side,lady,torio,torio2}
However, for $N=3,5,...$ the peak is split in two by a deep minimum. This result is in
contradiction with previous results using PTU which predict ideal conductance $G=G_0$
at the EHSC.\cite{oguri} Therefore, B{\"u}sser {\it et al.} conclude that other many-body
techniques should be used to elucidate the issue.

The purpose of the present work is to present an analytical approach
to calculate the conductance through an interacting region, weakly coupled
to conducting leads. As an application, we study the above mentioned controversy.
In particular, we consider an interacting region composed of a linear array
of an odd number $N$
QD's (extension to even $N$ and other geometry is straightforward),
connected to two
non-interacting leads by the same hopping $V$.
We also study the conductance for the case in which the interacting region
consists of an isosceles triangle of QD's.
To simplify the discussion
when the energy of states with different number of particles are compared,
we set the origin of
one-particle energies at the Fermi level $\epsilon_F=0$. In general, if the ground
state of the interacting part is quasi degenerate between a non-magnetic singlet and
a magnetic doublet, and $V$ is small compared with the
difference in energy with other states, neglecting the latter the problem can be mapped
into a one-impurity Anderson model.
As stated above, this model is well understood. This mapping has been
used to calculate the conductance through a ring described by the ionic Hubbard model,
which should detect a topological phase transition.\cite{ihm} In the Kondo regime, where
the magnetic ground state lies well below the other states, it might be necessary to retain
other states with even total spin for an accurate description.\cite{trimer}.
As in these problems, we assume that the ground state of the
interacting part is a doublet for an odd number of particles.
In particular, the Hubbard model used to represent a linear array
of $N$ QD's has a doublet ground state in the EHSC for odd $N$.
We retain this doublet and all singlet states which hybridize with it to map the problem
into a generalized one-impurity Anderson model. The validity of the mapping is discussed
in detail in section \ref{sec:summary}. The resulting model is solved using
a slave-boson representation in the saddle-point approximation. Explicit calculations are presented for
$N=1$, a linear trimer and an isosceles triangle. The main result is independent of
odd $N$ for a linear array of QD's: we obtain ideal conductance
in the EHSC. This agrees with recent results obtained using
alternative techniques.\cite{ogu3d,nisi,zit2}
The isosceles trimer is of interest because of the effects of magnetic frustration and peculiar electronic structure. \cite{trimer, jam, rusos1, rusos2, laza, inge}. In this case the electron-hole symmetry is lost and this fact is reflected in an asymmetric lineshape for the conductance. Despite this fact, the unitary limit is reached when $V_g$ favours a magnetic doublet ground-state.
 
The paper is organized as follows. In Sec. \ref{sec:model&app}, we describe the model,
the mapping to one effective site, and the slave-boson formalism for the
resulting effective model. The results for the conductance are contained
in Sec. \ref{sec:cond}. Sec. \ref{sec:summary} summarizes our results and discusses our findings in
relation to previous works. Some details of the calculations are left
to the appendices.

\section{Model and approximations}\label{sec:model&app}
\subsection{Model}\label{sec:model}
For the sake of clarity in the presentation, we consider a linear array of
an odd number $N$ of QD's coupled to the left
and to the right to metallic leads, and take $N=3$ unless otherwise stated.
The changes for any other array of QD's  are obvious. The system
is represented in Fig. \ref{sistema}. The dots are equivalent and their on-site
energy $E_d=-eV_g$ is controlled by the gate potential $V_g$. The leads
are described by non-interacting half-infinite chains.

\grafico{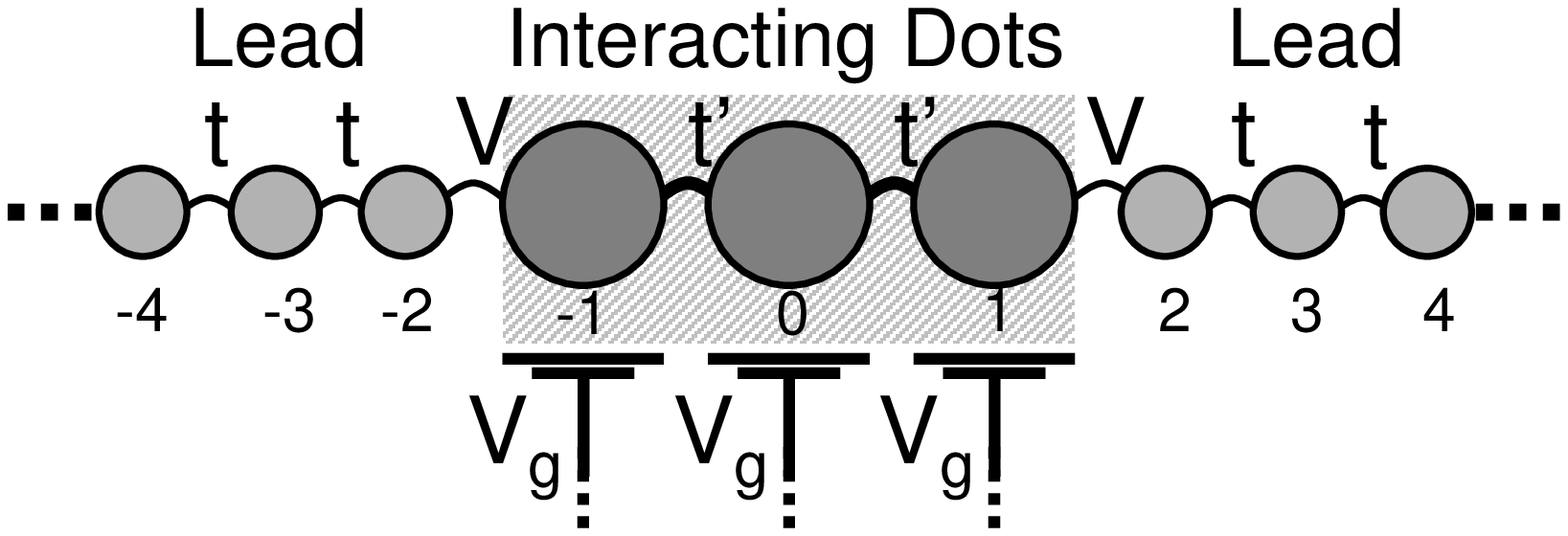}{0.5}{Scheme of the studied system.}{sistema}

The Hamiltonian is
\eqarray{
H&=&H_{\Rm{l}}+H_{\Rm{d}}+H_{\Rm{mix}}.
}{ham}
$H_\Rm{l}$ represents the non-interacting leads
\eqarray{
H_\Rm{l}&=& \sum_{i=-2,\sigma}^{-\infty}
\left( -t\ c^{\dagger}_{i,\sigma}c_{i-1,\sigma}+ \Rm{H.c.}\right) + \nonumber\\
&&\sum_{i=2,\sigma}^\infty \left(-t\ c^{\dagger}_{i,\sigma}c_{i+1,\sigma}+ \Rm{H.c.}\right).
}{hc}
$H_\Rm{d}$ describes the central region containing the dots,
each with an on-site repulsion $U$,
\eqarray{
H_\Rm{d}&=&\sum^{1}_{i=-1,\sigma} \left( E_d \ c^{\dagger}_{i,\sigma}c_{i,\sigma} + U\ n_{i,\uparrow}n_{i,\downarrow}\right) - \nonumber\\
&&\sum_\sigma \left[t'\ \left(c^{\dagger}_{-1,\sigma}c_{0,\sigma} +c^{\dagger}_{0,\sigma}c_{1,\sigma} \right) + \Rm{H.c.}\right].
}{hd}
Finally the term $H_{\Rm{mix}}$, that couples the interacting part of
the system with the conducting leads has the form
\eqarray{
H_{\Rm{mix}}&=& V \left[\sum_\sigma \left(c^{\dagger}_{-1,\sigma}c_{-2,\sigma}+ c^{\dagger}_{1,\sigma}c_{2,\sigma}\right)+\Rm{H.c.}\right].
}{hmix}
We take the Fermi energy $\epsilon_F$=0, in such a way that the leads are
half filled. If in addition $V_g$ is such that $E_d=-U/2$,
the Hamiltonian (for a linear array of QD's) is invariant
under an electron-hole transformation
$c^{\dagger}_{i,\sigma} \rightarrow (-1)^i c_{i,\sigma}$.
In addition,
for any value of $V_g$, the system is invariant under inversion
$c^{\dagger}_{i,\sigma} \rightarrow c^{\dagger}_{-i,\sigma}$ (for odd $N$).

\subsection{Mapping of the interacting region to an effective site}\label{sec:mapping}
We assume that the state of lowest energy of the interacting part
of the system for a certain odd number of particles of interest is
a doublet. This is certainly true for the ground state of Eq. (\ref{hd}) in the EHSC.
$H_\Rm{mix}$ mixes this doublet with other states with
$N-1$ and $N+1$ particles.
For small enough $V$, $H_\Rm{mix}$ can be eliminated trough a canonical
transformation, and the model can be mapped to a single impurity Kondo
system,\cite{trimer} which is well understood. However, as $V_g$ varies,
the state of minimum energy in the sector of either $N-1$ or $N+1$ particles
becomes quasi degenerate with the ground state and finally crosses
it, rendering the canonical transformation invalid. Near the
quasi degeneracy the model can be mapped into a one-impurity
Anderson model.\cite{ihm} Here we generalize previous approaches
\cite{ihm,trimer} and map the system into a generalized Anderson
model, retaining the lowest doublet in the subspace of $N$ particles,
and all singlet states in the subspaces with $N \pm 1$ (see Fig. \ref{subspaces}).
For simplicity we neglect the triplet states, but as we will show
this does not modify our conclusions. The neglect of excited
doublets is the most serious and its effect is discussed
in Sec. \ref{sec:summary}.

The eigenstates of $H_\Rm{d}$ are classified according to its
parity ($\nu=\pm$), total spin  $S$ and its projection $S_z$.
We will denote these eigenstates as: $|\psi^{(n)}_{j,\Bf{p}}\ra$,
where $n$ is the number of particles, $j$ orders the states
according to increasing energy and $\Bf{p}$ is a short label for
the other  quantum numbers ($\nu$, $S$ and $S_z$). The corresponding
energies are denoted by $E^{(n)}_{j,\Bf{p}}$.

\grafico{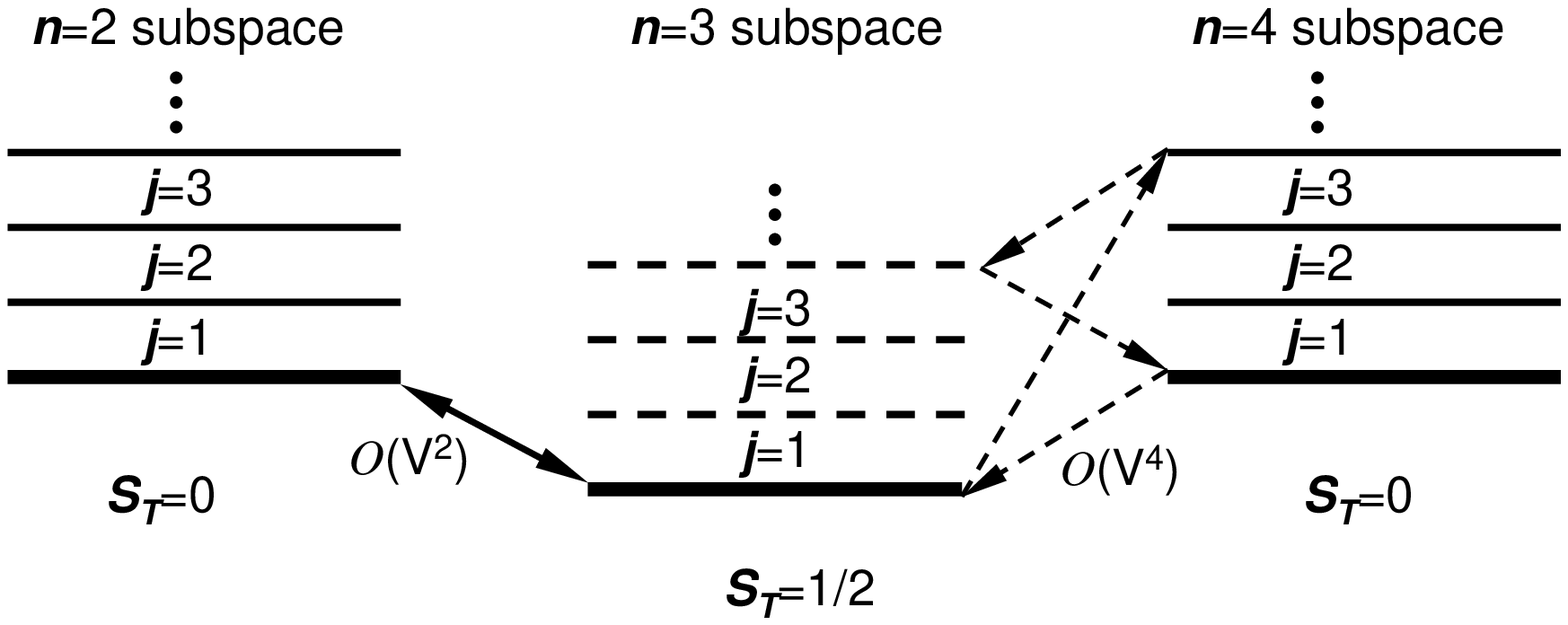}{0.4}{Scheme of the low-lying states
of $H_\Rm{d}$ for $N=3$. The dashed lines denote states truncated
in the approximation. Dashed arrows represent processes of order $V^4$, which are neglected. Only processes of order $V^2$ are retained.}{subspaces}

To represent the truncated Hamiltonian, we will use a slave boson
formalism. This is convenient in our case in which, in addition to the
truncation, we perform a saddle-point approximation to the slave boson
formalism. However, the slave boson representation is independent
of the method of solution chosen, and has been used for example in
exact diagonalization of finite clusters including triplet
states.\cite{simon}

\subsection{Formalism of slave bosons}\label{sec:sbformalism}
As stated before, for the sake of clarity we will refer to the case of the linear chain. We represent the many-body states of $H_\Rm{d}$ using a slave boson
representation that respects electron-hole symmetry. This symmetry is important for the description of the conductance through a linear array, as discussed in the introduction. Basically,
this representation can be described as a generalization
of that used by Kotliar and Ruckenstein for the Hubbard model \cite{kr} to
include several ``empty" $e$ and ``doubly occupied" $d$ bosons.
Specifically, we introduce the boson operators $e_{j,\nu}, d_{j,\nu}$ and $s_{\sigma}$
and the fermionic ones  $f_\sigma$, to map the states of the
effective site describing the interacting part as
\eqarray{
|\psi^{(2)}_{j,\nu}\ra&\rightarrow& e^{\dagger}_{j,\nu}|0\ra, \nonumber \\
|\psi^{(3)}_{0,-,\sigma}\ra&\rightarrow& f^{\dagger}_{\sigma}s^{\dagger}_{\sigma}|0\ra, \nonumber \\
|\psi^{(4)}_{j,\nu}\ra&\rightarrow& d^{\dagger}_{j,\nu}f^{\dagger}_{\uparrow} f^{\dagger}_{\downarrow}|0\ra,
}{repbosones}

with the following constraints
\eqarray{
\sum_{j,\nu} e^{\dagger}_{j,\nu}e_{j,\nu}+\sum_{\sigma} s^{\dagger}_\sigma s_\sigma+\sum_{j,\nu} d^{\dagger}_{j,\nu}d_{j,\nu }&=&\mathbf{1},
}{constraint1}
\eqarray{
f^{\dagger}_{\sigma}f_{\sigma}&=&s^{\dagger}_{\sigma}s_{\sigma}+\sum_{j,\nu} d^{\dagger}_{j,\nu}d_{j,\nu }.
}{constraint2}

We point out here that for a linear trimer, the ground state
$|\psi^{(3)}_{0,-,\sigma}\ra$
of the interacting part $H_\Rm{d}$ is odd under inversion.
This implies that one has to choose one of the operators
$f_\sigma$, $s_\sigma$ even and the other odd under
inversion. Otherwise the invariance under inversion would
be violated.

In this representation, the Hamiltonian can be written as
\eqarray{
H&=&H_\Rm{l}+H_\Rm{f}+H_\Rm{mix}+H_\Rm{boson}+H_\Rm{constr},
}{ham_boson}
where $H_\Rm{l}$ is the same as before (Eq. (2)).  $H_\Rm{f}$ describes the energy of the
fermions at the effective site
\eqarray{
H_\Rm{f}&=& E^{(3)} \sum_\sigma f^{\dagger}_\sigma f_\sigma,
}{hf}
where $E^{(3)}\equiv E^{(3)}_{0,-,\sigma}$ is the energy of the lowest doublet
in the subspace of three particles (we keep $N=3$ in this Section for
simplicity).
Similarly for the bosons at the effective site
\eqarray{
H_\Rm{boson}&=&\sum_{j,\nu}
E^{(2)}_{j,\nu}\ e^{\dagger}_{j,\nu}e_{j,\nu}+ \sum_{j,\nu} \left[
E^{(4)}_{j,\nu}- 2 E^{(3)} \right] \ d^{\dagger}_{j,\nu}d_{j,\nu}.\nonumber\\
}{hboson}
The energy of the boson  $s_\sigma$ has been chosen at zero.
Note that the above choice ensures the correct energy for each state
of the effective site:
\eqarray{
P_0 H P_0 \ e^{\dagger}_{j,\nu}|0\ra &=& E^{(2)}_{j,\nu} \ e^{\dagger}_{j,\nu}|0\ra, \nonumber \\
P_1 H P_1 \ f^{\dagger}_{\sigma}s^{\dagger}_{\sigma}|0\ra &=& E^{(3)} \  f^{\dagger}_{\sigma}s^{\dagger}_{\sigma}|0\ra, \nonumber\\
P_2 H P_2 \ d^{\dagger}_{j,\nu}f^{\dagger}_{\uparrow} f^{\dagger}_{\downarrow}|0\ra&=& \left[ (E^{(4)}_{j,\nu}-2 E^{(3)})+2 E^{(3)}\right]\times\nonumber\\
&& \times d^{\dagger}_{j,\nu}f^{\dagger}_{\uparrow} f^{\dagger}_{\downarrow}|0\ra,
}{energia_conf}
where we have used the projectors $P_i$ over the subspace with $i$
fermions, or $n=i+2$ ($n=i-1+N$ in general) particles in the interacting
region.

Defining for convenience electron operators with well
defined parity under inversion
\eq{
c_{|i|,\pm,\sigma}=\frac{c_{i,\sigma}\pm c_{-i,\sigma}}{\sqrt{2}},
}{simops}
the term $H_\Rm{mix}$ in this representation becomes
\begin{widetext}
\eqarray{
H_\Rm{mix}&=& \sqrt{2}V \left\{\sum_{2,\nu,\sigma} f^{\dagger}_{\sigma}c_{-\nu,\sigma}\left[ s^{\dagger}_{\sigma} \left(\sum_{j} \alpha_{j,\nu} e_{j,\nu}\right) + \left(\sum_{j} \beta_{j,\nu} d^{\dagger}_{j,\nu}\right) s_{\overline{\sigma}} \right] + \Rm{H.c.}\right\},
}{hmixboson}
\end{widetext}
where
\eqarray{
\alpha_{j,\nu}&=&\la \psi^{(3)}_{0,\nu_g,\sigma}|c^\dagger_{1,(\nu . \nu_g),\sigma}|\psi^{(2)}_{j,\nu}\ra,\\
\beta_{j,\nu}&=&\la \psi^{(4)}_{j,\nu}|c^\dagger_{1,(\nu . \nu_g),\sigma}|\psi^{(3)}_{0,\nu_g,\overline{\sigma}}\ra,
}{alphabeta}
where $\nu_g$ is the parity of the ground-state for $N$ particles. For the linear array, $\alpha_{j,\nu}$ and $\beta_{j,\nu}$takes the same values as a consequence of the electron-hole symmetry for a particular $V_g$
(as mentioned above) and the fact that both matrix elements turn out to be independent of $V_g$.

The constraints are incorporated in the Hamiltonian introducing
Lagrange multipliers $\lambda '$ and $\lambda_\sigma$, corresponding to
the Eqs.  (\ref{constraint1}) and Eq. (\ref{constraint2})
\begin{widetext}
\eqarray{
H_\Rm{constr}&=&\lambda' \left( \sum_{j,\nu}e^{\dagger}_{j,\nu}e_{j,\nu} + \sum_{\sigma}s^{\dagger}_{\sigma}s_{\sigma} + \sum_{j,\nu}d^{\dagger}_{j,\nu}d_{j,\nu} - \Bf{1} \right) + \sum_{\sigma} \lambda_{\sigma} \left( f^{\dagger}_{\sigma}f_{\sigma} - s^{\dagger}_{\sigma}s_{\sigma} - \sum_{j,\nu}d^{\dagger}_{j,\nu}d_{j,\nu} \right).
}{hconstr}

In the functional integral formalism, the fermions can be integrated out
as in Hubbard model,\cite{kr} and the partition function becomes
\eqarray{
Z&=&\int \prod_{j,\nu}[{\cal D} e_{j,\nu}]\prod_{j,\nu}[{\cal D} d_{j,\nu}]\prod_{\sigma}[{\cal D} s_{\sigma}]d\lambda_{\sigma} d\lambda' \ \exp{\left[-\int_0^{\beta} d\tau \ \tilde{S}(\tau)\right]},
}{Z}
where the effective action is
\eqarray{
\tilde{S}(\tau)&=&\sum_{j,\nu} \left[e^{*}_{j,\nu}\ (\partial_{\tau}+E^{(2)}_{j,\nu}+\lambda')\ e_{j,\nu} \right]+
\sum_{\sigma} \left[s^{*}_{\sigma}\ (\partial_{\tau}+E^{(3)}+\lambda'-\lambda_{\sigma})\ s_{\sigma} \right]+\nonumber\\
&&+ \sum_{j,\nu} \left[d^{*}_{j,\nu}\ (\partial_{\tau}+E^{(4)}_{j,\nu}-2E^{(3)}+\lambda'-\sum_{\sigma} \lambda_{\sigma})\ d_{j,\nu} \right]-
\lambda'+\sum_{\nu,\sigma} \Rm{Tr} \ \ln{M_{\nu,\sigma}},
}{Seff}
and the non-zero matrix elements $M_{\nu,\sigma}$ (for $N=3$) are
\[
\left[M_{\nu,\sigma}\right]_{i,l}=\left\{
\begin{array}{lll}
\left( \partial_\tau -\mu \right) \delta_{i,l}-t \left( \delta_{i,l+1}+\delta_{l+1,j} \right)&\mbox{if}& i,l \ge 2,\\
\left( \partial_\tau -\mu +E^{(3)}-\lambda_\sigma \right)& \mbox{if}&i=l=0,\\
V_{\nu,\sigma}&\mbox{if}&(i=0;l=2)\ \mbox{or}\ (i=2;l=0)
\end{array}\right.,
\]
\end{widetext}
where
\eqarray{
V_{\nu,\sigma}&=&\sqrt{2} V \left[ s^{*}_{\sigma} \left(\sum_{j} \alpha_{j,\nu} e_{j,\nu}\right) +
\left(\sum_{j} \beta_{j,\nu} d^{*}_{j,\nu}\right) s_{\overline{\sigma}} \right].\nonumber\\
}{Vnu}

Up to now, the only approximation made was the truncation for $N >1$ of the
Hilbert space, to a set of relevant low-energy states of the interacting part
$H_\Rm{d}$. For $N=1$, the formalism introduced in this subsection
is just a change of representation of the original Hamiltonian.

To solve the problem we used the saddle-point approximation to evaluate
the partition function. The bosonic fields are replaced by real constant
numbers that minimize the action. The details are left for
appendix \ref{sec:spa}.

\section{Results for the conductance}\label{sec:cond}
In the slave-boson mean-field approximation that we are using,
the many-body problem is reduced to a non-interacting one with
one ``impurity" and only one channel of conduction, with  renormalized
hybridization (see Eq. (\ref{Vnu})). Then we can use the two-terminal
Landauer equation \cite{meir} obtaining
\eqarray{
G=\frac{2e^2}{h}\int d\omega \left(-\frac{\partial  f(\omega)}{\partial \omega}\right) \left|\Bf{t}(\omega)\right|^2,
}{landauer}
where $f(\omega)$ is the Fermi function and $\Bf{t}(\omega)$ is the transmittance through the effective site
\eqarray{
\Bf{t}(\omega)&=&\pi \rho_0 \left(V_{-2}V_{2}\right) G_{ff \sigma}(\omega)\nonumber\\
&=&2 \pi \rho_0 \left(V^2_+ - V^2_-\right) G_{ff \sigma}(\omega),
}{t}
$G_{ff \sigma}(\omega)$ is the retarded Green's function of
the fermion $f$ with spin $\sigma$ (see appendix \ref{sec:spa}),
and $V_{\pm 2}=\left(V_{+}\pm V_{-}\right)/\sqrt{2}$ is the coupling between the effective site
and (for $N=3$) the site with subscript $\pm 2$.
At zero temperature from the above equations one gets
\eqarray{
G&=&\frac{2e^2}{h} \left[2 \pi \rho_0 \left(V^2_+ - V^2_- \right) \right]^2  |G_{ff \sigma}|^2_{\omega=0}.
}{landauer2}

\subsection{Conductance for $N=1$}
We first discuss the case of the conductance through one QD, and we compare
the results with those obtained with NRG \cite{nrg1} for different values of
$\Delta $, which is the half width at half
maximum of the resonant level in the non-interacting case.
In our problem $\Delta =2\pi \rho
_{0}V^{2}$, where $\rho _{0}=1/(2 \pi t)$ is the unperturbed density of states
at the Fermi energy for one
spin and one of the leads, and the factor 2 adds the contribution of both
leads.

The results for one dot are shown in Fig. \ref{comp1d}. The ideal conductance
for $E_d=-U/2$ indicates the formation of a virtual bound state (Kondo
effect) at $T=0$. As can be seen from the inversion symmetry of the figure
around $E_{d}=-U/2$, the solutions of the saddle-point equations reflect the
electron-hole symmetry of the formalism. A very good agreement between our
results and those of NRG for $\Delta /\pi U=0.01$  (strongly-interacting
case) is observed. This is in part due to the fact that in the EHSC, where
analytical results can be obtained (see appendix \ref{sec:verification_fsr}),
the formalism
reproduces the correct exponential dependence of the width of the resonance
in the spectral density of states $\delta$, which roughly coincides with
the Kondo temperature $T_{K}$. However, the shape of the resonance is
Lorentzian in the mean field approximation, while the correct result in the
Kondo regime is \cite{frota}

\begin{equation}
\rho _{d\sigma }(\omega )=\frac{1}{\pi \Delta } {\rm Re}\sqrt{(\omega
+i\delta )/i\delta },  \label{rhonrg}
\end{equation}
where $\delta \cong 1.55T_{K}$.\cite
{campo}

The good agreement between the slave-boson mean-field approximation and NRG deteriorates
as smaller values of $U$ are considered and is lost in the
non-interacting limit. Kotliar and Ruckenstein proposed a remedy to this problem,
replacing some operators entering  $H_\Rm{mix}$ in Eq.(\ref{hmixboson})
by another ones (containing some suitable chosen roots) which
coincide with the previous ones when the constraints are imposed exactly on
each site, and at the same time reproduce in the saddle-point approximation,
the correct results for the noninteracting case.\cite{kr}
Unfortunately, we find that this method does not work in the whole
range of $\Delta/\pi U$, giving an overestimated Kondo temperature in
the strongly-interacting limit. In contrast, as mentioned above, the formalism
so far presented gives the correct dependence of $T_K$ on the parameters
in this limit (see appendix \ref{sec:verification_fsr}).

\grafico{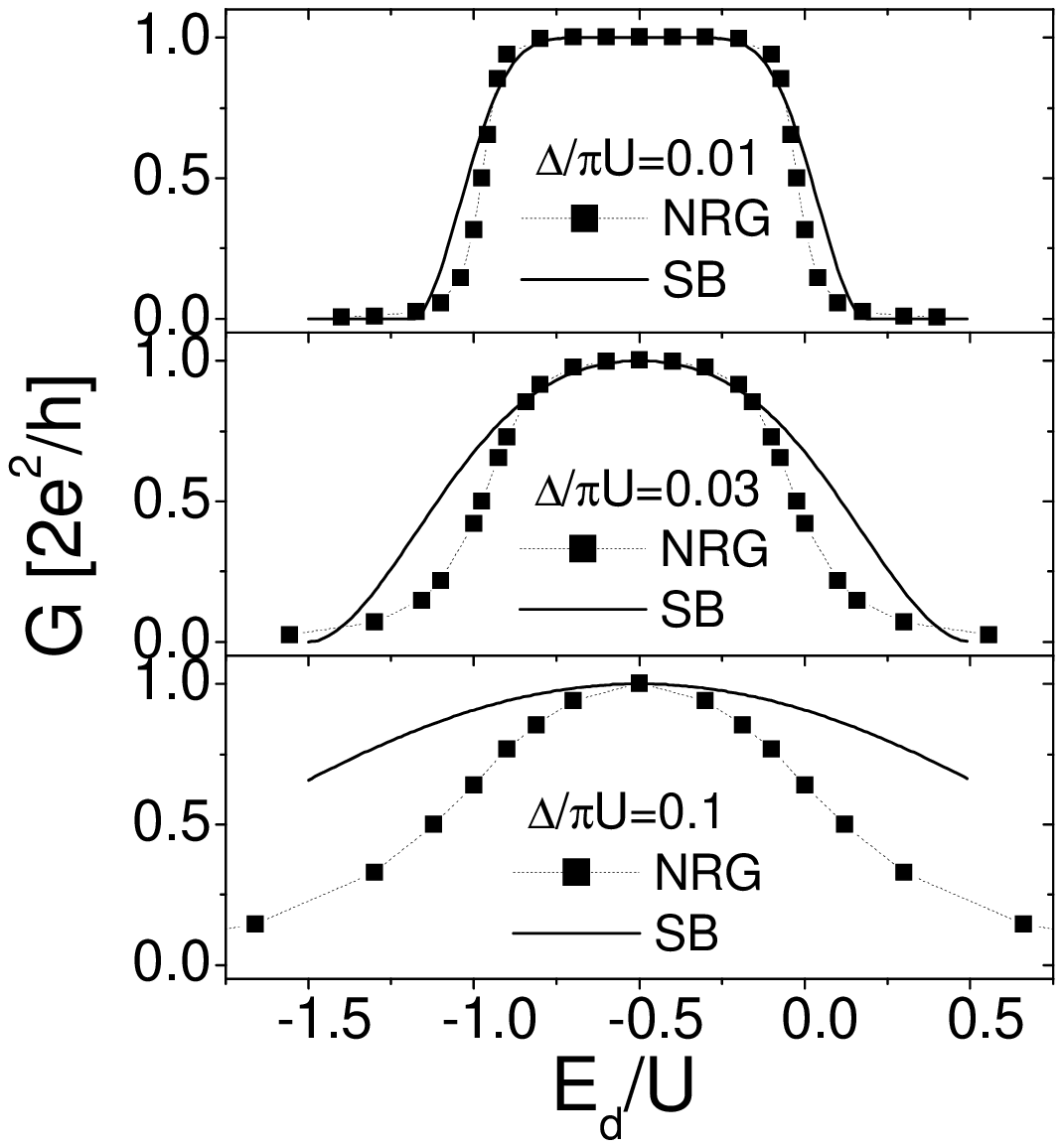}{0.6}{Conductance vs $E_d/U$ ratio, for different
values of $\Delta/\pi U$. Lines indicate the results of the present slave boson (SB) approach, and the square points have been obtained with
NRG. \cite{nrg1}}{comp1d}

Since we are mainly interested in the limit of strong correlations,
the results shown in Fig.   \ref{comp1d} are encouraging.
For small or moderate values of $U$ one can use PTU.\cite{side,lady,oguri,he2}
In spite of this good agreement, the conductance falls to zero abruptly
at some value of $E_d$. This is an artifact of the
saddle-point approximation, reported before in other problems,\cite{ihm,lady,hew}
for example in the conductance through rings of quantum dots.\cite{ihm} Beyond some value
of the gate voltage $V_g$, the impurity decouples from the rest of the chain.

The decrease in the conductance out of the EHSC is due to the
reduction of the density of states at the Fermi level, as the system
approaches the intermediate valence regime. Technically,
in the saddle-point approximation as $E_d$ increases,
the magnitude of boson $e$ also increases (the charge in the dot decreases).
This causes the magnitude of the boson $s$ to decrease and the resonance
in $f$-electron density of states must be shifted to higher frequencies
in order to decrease the $n_f$ occupation, according to the constraint
Eq. (\ref{constraint2}). This causes a reduction in the value of $|G_{ff \sigma}|$
at the Fermi level, and a reduction in the conductance
(see Eq. (\ref{landauer})). In a similar fashion, when the dot is charged,
the value of the boson $d$ increases, the value of $n_f$ approaches to $1$,
and the resonance in the density of states must be shifted to lower
frequencies, producing an analogous decrease in the conductance.

For the simple case of conduction through one dot, the expression of the
conductance can be written as \cite{meir,side}
\eqarray{
G&=&\frac{2e^2}{h} \int d\omega \ \left(-\frac{\partial f}
{\partial \omega}\right) \pi \Delta \ \rho_{d\sigma }(\omega) \ ,
}{cond1d}
At very low temperatures ($T\ll T_{K}$), the system is a Fermi liquid.
Therefore, the Friedel's sum rule (FSR) \cite{lang} must hold. For the case
in which the hybridization $V$ and the unperturbed density of
conduction states $\rho _{0}$ do not depend on energy, the FSR takes the
form
\begin{equation}
\rho _{d\sigma }(0)=\frac{\sin ^{2}(\pi \langle n_{d\sigma }\rangle )}{\pi
\Delta },  \label{fsra}
\end{equation}
giving a conductance equal to

\begin{equation}
G=\frac{2e^{2}}{h}\sin ^{2}(\pi \langle n_{d\sigma }\rangle ){,}
\label{fsrc}
\end{equation}
where $\langle n_{d\sigma }\rangle $
is the total charge per spin in the dot.

It can be shown (see appendix \ref{sec:verification_fsr}) that the
saddle-point approximation verifies the FSR, since the real problem is mapped into an effective
non-interacting one whose ground state is a Fermi sea.

\subsection{Conductance for $N=3$. Linear trimer}
The results for conductance through three QD's using the saddle-point
approximation described in appendix \ref{sec:spa}, are shown in
Fig. \ref{3d1d}.
Results for one QD for the same parameters are also shown for comparison.

\grafico{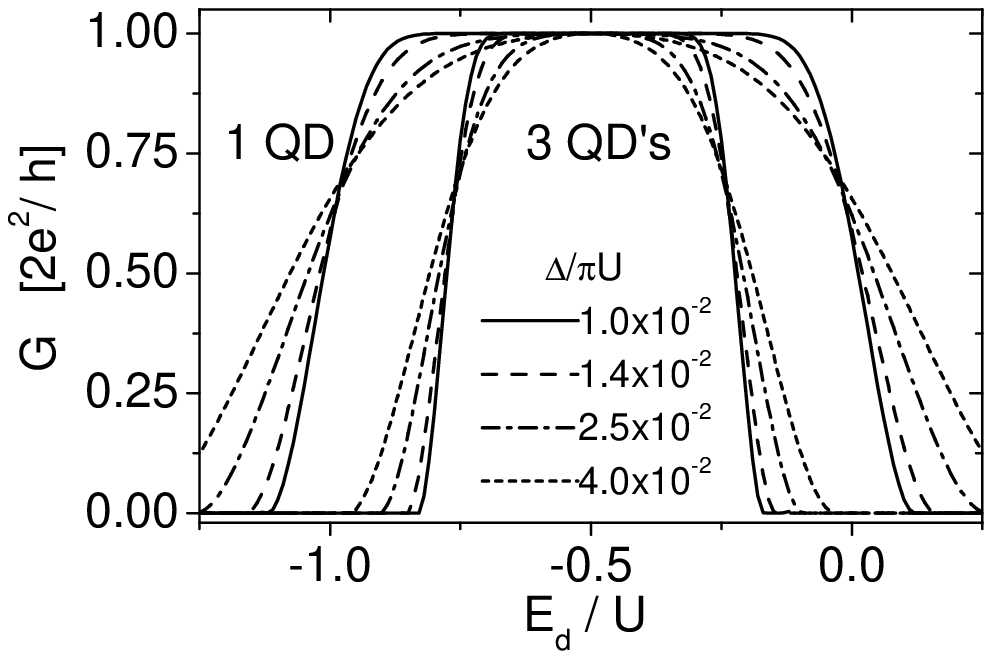}{0.75}{Conductance vs $V_g/U$ ratio, for different
values of $\Delta/\pi U$. Innermost curves correspond to the conductance through 3 QD's. The 1 QD conductance is also shown for comparison. Parameters are $U=t$, $t'=0.5\ t$ and the hybridization $V$ is 0.25 $t$, 0.3 $t$, 0.4 $t$ and 0.5 $t$ for increasing values of $\Delta/\pi U$.}{3d1d}

On a qualitative level, the results for three QD's are similar to those of
one QD near the EHSC. In particular, the conductance equals the ideal one $G_0$ in the EHSC,
decreases as the amount of charge fluctuations increases, and finally
becomes very small when the ground state of the interacting region
becomes non-magnetic.
The reduction of the region in which the conductance is near to the ideal one
is reduced in the three-dot case, and one expects further reduction as the
odd number of dots $N$ increases. This is due to the fact that as $N$
increases, the energy of singlets with $N \pm 1$ particles become nearer
to that of the ground-state doublet with $N$ particles. As a consequence,
the interval of gate voltage $V_g$ between both intermediate valence regimes
(when the energy of one singlet coincides with that of the doublet) is
reduced.

When $V_g$ is moved further from the EHSC, an important difference between
$N=1$ and odd $N>1$ is that in the latter case, magnetic states with
$N \pm 2$ particles become accessible, a new Kondo effect takes place,
and another plateau with near ideal conductivity appears. For $N>3$
this pattern with regions of nearly ideal or zero conductivity can be
repeated several times. In general, including even $N$, there should be
plateaus when the gate voltage favors an odd number of particles.
In particular, for $N=2$ the conductance in a similar system has been
calculated using NRG and displays two plateaus, corresponding to 1 and 3
particles in the interacting region.\cite{zit}. While we have limited
our calculations to odd $N$ and gate voltages such that either $N$ or
$N \pm 1$ particles are favored, it is straightforward to extend the
formalism to other cases, as long as the hybridization $V$ of the interacting
region with the rest of the system is small enough. The allowed magnitude
of $V$ for the validity of the formalism is discussed in the next section.

We have verified numerically that Eq. (\ref{fsrc}) also holds for three dots.
The formalism is identical for other odd $N$.
Therefore, near the EHSC the situation for different odd $N$
are qualitatively similar in that they show a
local Fermi liquid behavior, ideal conductance at the EHSC, and
the expected electron-hole symmetry. This is at variance
with the results of B\"usser {\it et al.} \cite{busser} who
find vanishing conductance at the EHSC, but are consistent with
those of Oguri,\cite{oguri} and recent studies for $N=3$,\cite{ogu3d,zit2}
who also obtain ideal conductance at
the EHSC. Note that the parameters with $V=0.3\ t$ used in Fig. \ref{3d1d}
correspond to one set of parameters used in Fig. 2 (f) of
Ref. \onlinecite{busser}.

\subsection{Conductance for $N=3$. Isosceles triangle}

In this subsection, we consider a system with
the geommetry schematically depicted if Fig. \ref{isosc}.
Two regimes in parameter space were analized, namely $t'' < t'$
and $t'' > t'$, where $t''$ is the new hopping term connecting
dot 1 and -1. These regimes correspond to the conductance through
the two possible ``isosceles triangles''.
In the case $t'' = t'$ (equilateral triangle), the subspace with
$n=3$ has a doubly degenerate ground-state, with states belonging
to the even and odd subspace respectively. This can be understood
by recalling that the symmetry group of the equilateral triangle
in two dimensions ($C_{3v}$) has one two-dimensional irreducible
representation. As it will be discussed in detail in Section
\ref{sec:summary}, the present formalism is not valid at this
point of parameter space, since we assumed a
non-degenerate ground-state.
Since experimentally is very hard to reach such a regime, we believe
that this is not a serious constraint of the method. Starting with
the linear chain ($t''=0$), we recall that the ground-state of the
$n=3$ subspace is an odd $S=1/2$ doublet. Switching on adiabatically
the hopping $t''$, the energy of the excited even doublet begins to fall, and after
crossing the degeneracy point $t''=t'$, the even doublet takes over
and the formalism is valid again.

Besides these features, we expect the development of a plateau in
the conductance at the ideal value in a certain region of $E_d/U$,
due to the fact that the problem still can be mapped into an
effective $S=1/2$ Anderson impurity coupled to conducting leads.

\grafico{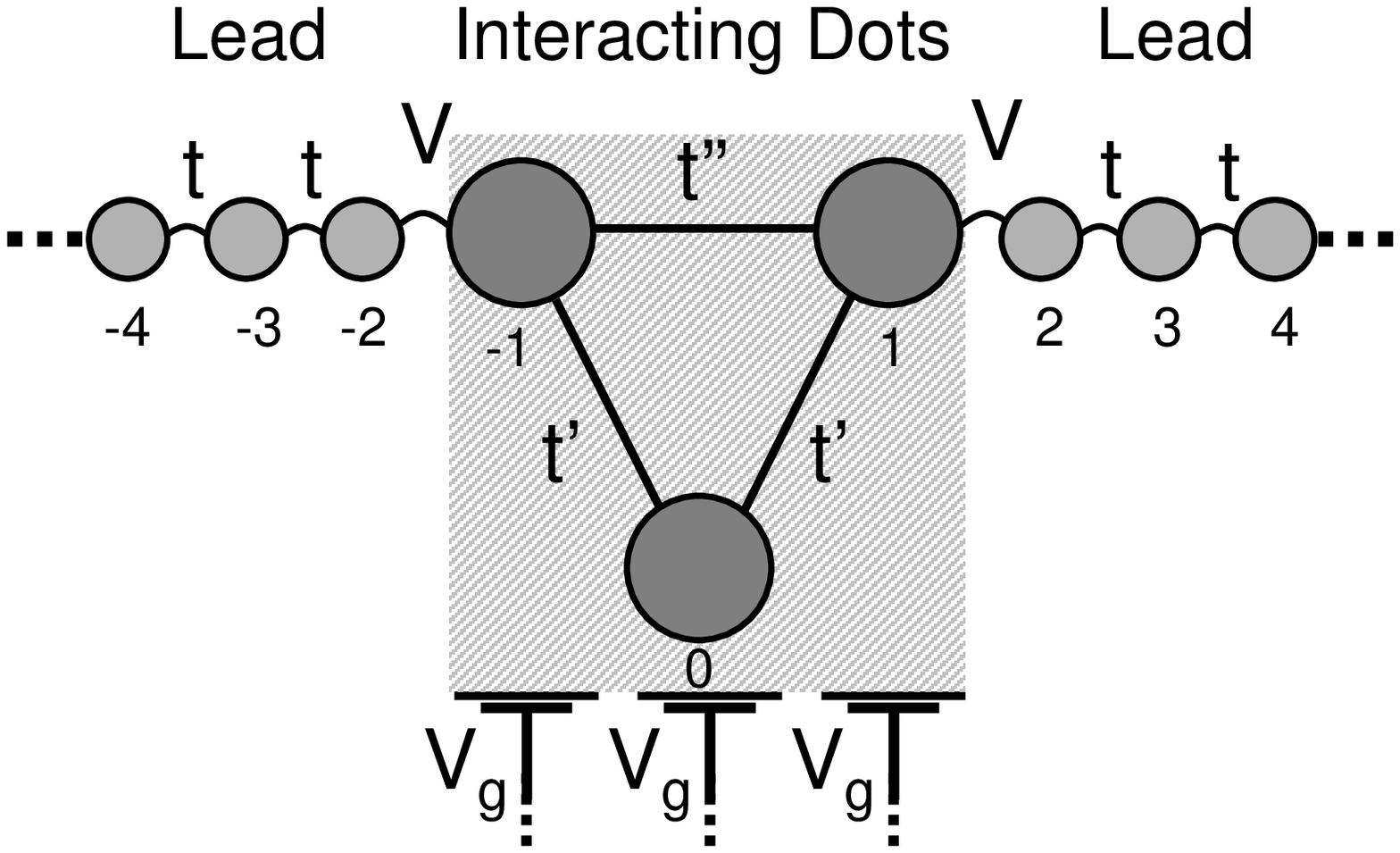}{0.45}{Scheme of the triangular arrangement.}{isosc}

In Fig. \ref{cond_isosc1} we show the results for the conductance through
the isosceles triangle. For simplicity we have set parameters
$V=0.45\ t, t'=0.5\ t$ and $U=t$, so that the first result ($t''=0$)
is one of the curves of Fig. \ref{3d1d}. As $t''$ is increased,
the curves are approximately rigidly shifted to lower values of
$E_d/U$. This feature can be correlated with properties of the
isolated trimer. In order to gain physical insight, we define the
following functions of the isolated trimer:

\begin{eqnarray}
\Delta_{23}(E_d,t'',t',U)&=&E^{(2)}_g(E_d,t'',t',U)-E^{(3)}_g(E_d,t'',t',U),\nonumber\\
\Delta_{43}(E_d,t'',t',U)&=&E^{(4)}_g(E_d,t'',t',U)-E^{(3)}_g(E_d,t'',t',U),\nonumber
\end{eqnarray}
\grafico{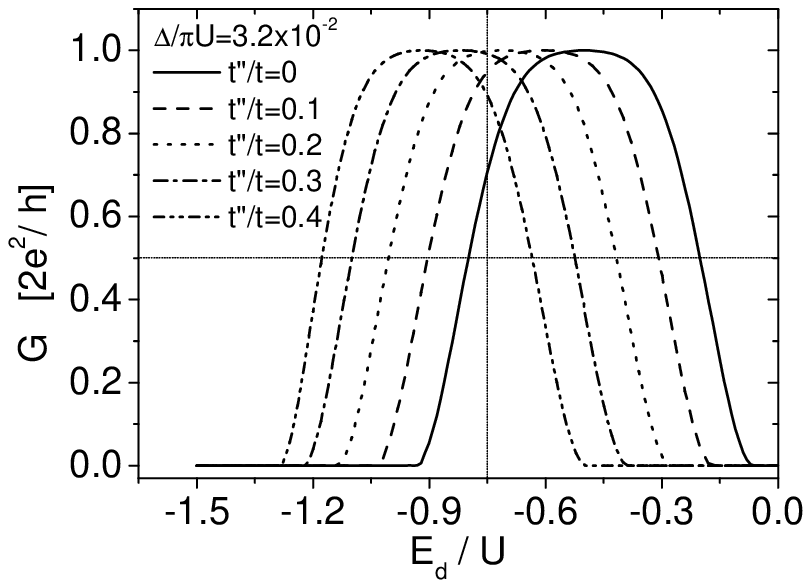}{0.9}{Results for the conductance through the isosceles triangle for the case $t''<t'$. Parameters are $t'=0.5\ t$, $U=t$ and $V=0.45\ t$.}{cond_isosc1}
which are the differences in energy between ground-states belonging to
subspaces with different particle number. Fixing $t'',t'$ and $U$,
we define $E^{2,3}_d$ and $E^{4,3}_d$ as the values of $E_d$ which render $\Delta_{23}(E_d,t'',t',U)=0$
and $\Delta_{43}(E_d,t'',t',U)=0$ respectively. These values correspond to an
intermediate valence regime of the efective Anderson model,
and are related to the crossover regions in the dependence of the
conductance $G$ with gate voltage, where $G$ falls down from its ideal
value at the plateau to zero. This is due to charge fluctuations and
the consequent dissapearance of the Kondo effect. Following the
evolution of the points $E^{2,3}_d$ and $E^{4,3}_d$ as a function of $t''$ (see Fig. \ref{dif_i}), the evolution
of the conductance curves in Fig. \ref{cond_isosc1} can be explained
easily: the conductance is related to the low-energy physics of the
isolated trimer. For example, for $t''=0\ (t''= 0.4\ t)$ , the first value of $E_d/U$ for which $G=G_0/2$ in Fig. \ref{cond_isosc1} is near $-0.8\ (-1.2)$, the value for which the number of particles in the ground-state of the isolated trimer changes from 4 to 3 (see Fig. \ref{dif_i}). As it is already clear from Fig. \ref{3d1d}, the width
of the crossover regions increases with $V$, since the effective
hybridization $V_{\nu,\sigma}$ is proportional to it and therefore the widths
of intermediate valence regimes increase accordingly.
\grafico{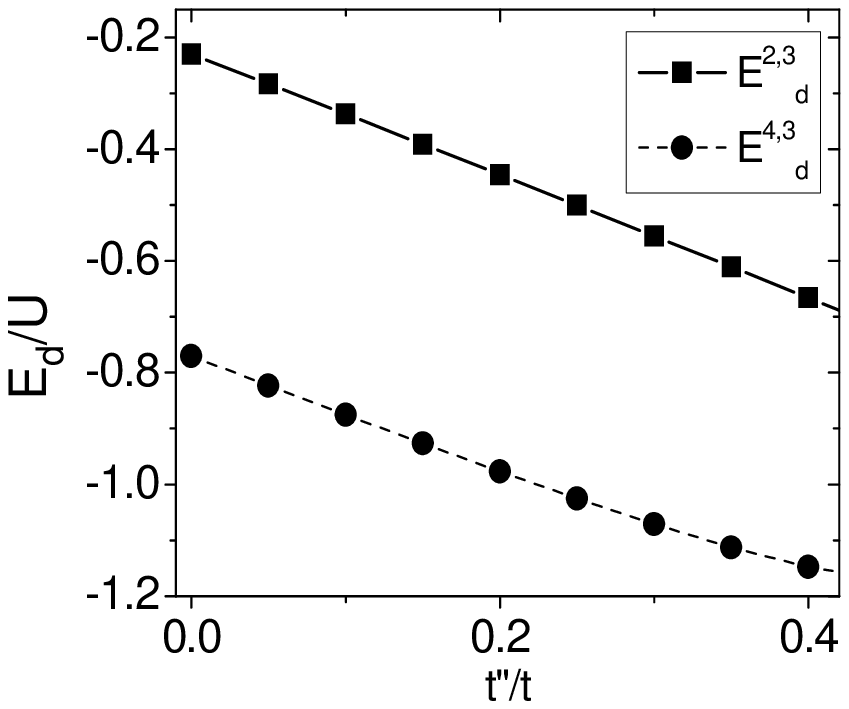}{0.8}{Evolution of the degeneracy points as a function of $t''$ for the isolated trimer. Parameters $t$ and $U$ are those of Fig. \ref{cond_isosc1}.}{dif_i}

\grafico{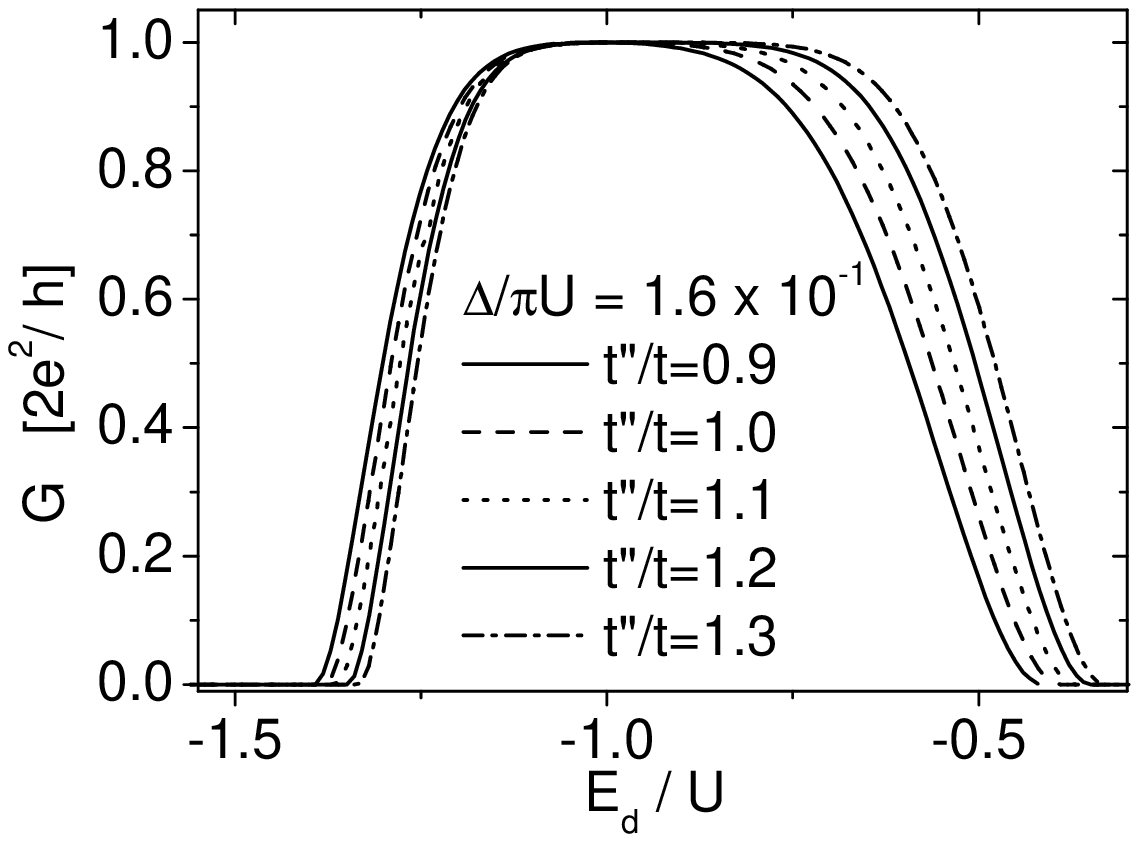}{0.65}{Results for the conductance through the isosceles triangle for the case $t''>t'$. Parameters are the same than Fig. \ref{cond_isosc1}.}{cond_isosc2}

\grafico{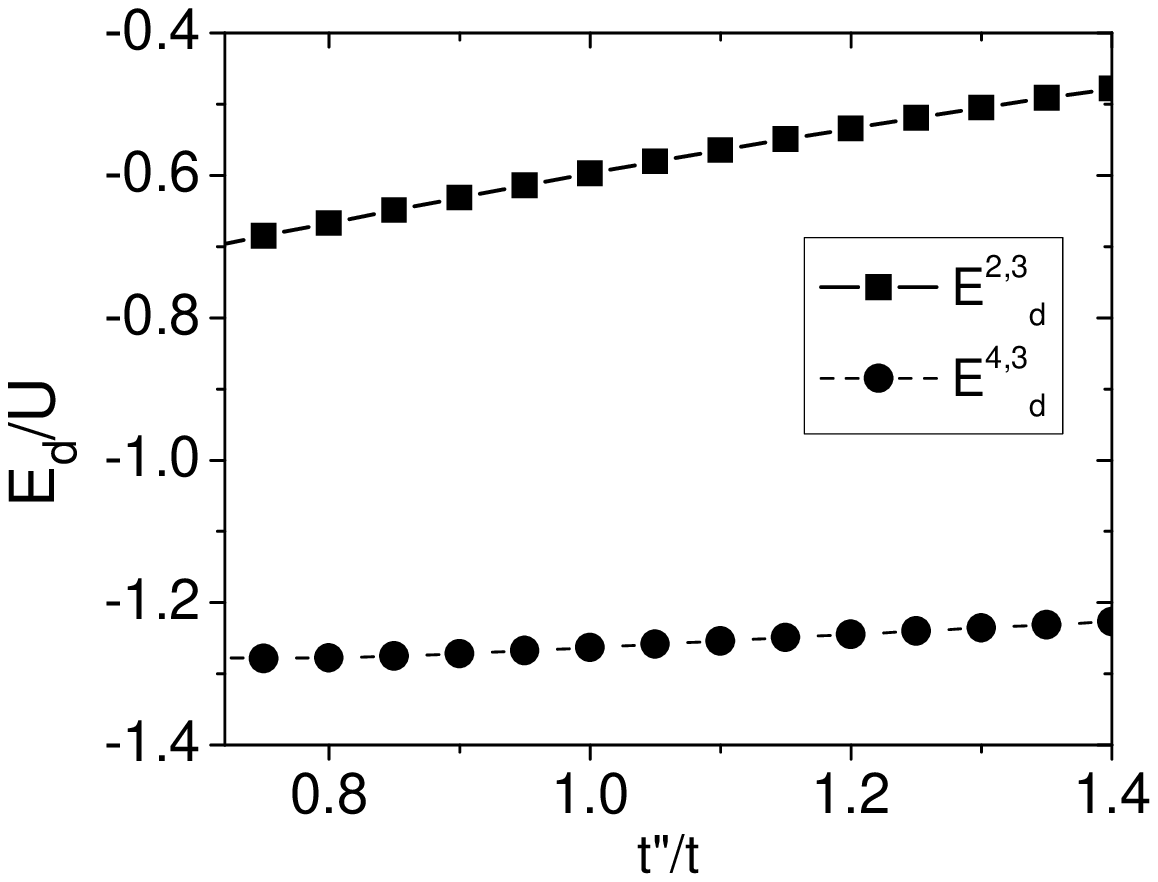}{0.65}{Same than Fig. \ref{dif_i} for the case $t''>t'$. The different evolution of the degeneracy points as a function of $t"$ explains the asymmetry of the conductance in Fig. \ref{cond_isosc2}.}{dif_p}

The results for $t''>t'$ are shown in Fig. \ref{cond_isosc2}. The asymmetry
in the curves is more evident, and again the plot can be understood
by following the evolution of $E_d^{2,3}$ and $E_d^{4,3}$ (see Fig. \ref{dif_p}).
When $t''>t'$, the low-energy physics of the system resembles
that of a singlet formed by the two QD's connected by $t''$ plus
an additional QD on the top. In this situation, there is a
considerable decrease in the effective hybridization
$V_{\nu,\sigma}$ (see Eq. \ref{Vnu}). This is directly reflected
in that the Kondo temperature is much lower than before,
rendering the convergence of the numerical method difficult.
For that reason, in those calculations we have set $V=1.0\ t$
in order to have a larger Kondo temperature. The validity of the
formalism is only slightly affected, despite the fact that $V$ is
comparable to other energy scales ($t,U$). The main reason for this
is the high values of $t''$, which ensure a large difference
between the ground state energy and that of the excited odd doublet
for $n=3$. Another reason is the small effective hybridization.
Other details on the validity of the approximation will be given in
the next section. As a final comment, we mention that the results obtained for these geommetries are consistent with 
those obtained with the FSR (Eq. (\ref{fsrc})) for the effective Anderson model.

\section{Discussion}\label{sec:summary}

We have presented an effective site approach to study the conductance
through an arrangement of $N$ QD's, diagonalizing exactly the
part of the Hamiltonian describing the interacting region and retaining
the most relevant states which describe the low-energy charge excitations.
As discussed below, the approach is valid when the hybridization $V$
of the interacting region with the rest of the system is small.
The advantage of this mapping is that the physics of one impurity
is rather well understood, and several good approximations can be used.
Recently a mapping to an effective Kondo model with one ``impurity" has
been successfully used to interpret the dependence of geometry
of spectroscopic experiments for a Cr trimer on Au(111),\cite{trimer}
and to calculate the conductance through a dimer connected at one site
to two conducting leads.\cite{zit} In the latter case, NRG results are available
and are in excellent agreement those of the effective model
for small or moderate $V$.\cite{zit}

For small $V$, the effective site approach can also be used when the
the interacting region is side coupled to a conducting lead and the
current does not necessarily flow through the interacting region. An
example of this system is a trimer coupled to one site of a conducting
lead.\cite{thp} In this case, the model is mapped to an effective
site side coupled to a conducting lead,\cite{side,lady,torio,torio2}
instead of embedded in the lead.

We have introduced a slave boson representation for the description
of the effective Hamiltonian, and used the saddle-point approximation
to calculate the conductance through the effective site.
The results obtained in the strongly-interacting regime for $N=1$
were found to be in very good agreement with NRG calculations.
For a linear array of an odd number of QD's,
our results near the EHSC are qualitatively similar. In particular,
we obtain perfect conductance at the EHSC. Since this result might seem
controversial at present, we discuss in detail the approximations
involved in our treatment, particularly in relation to this result.

The crucial approximation is to retain only the
lowest doublet of the magnetic configuration, neglecting the excited
ones (see Fig. \ref{subspaces}). This approximation scheme is only valid as far as
the excited levels have a negligible contribution to the many-body
ground state. The hybridization $V$ induces a second-order matrix element
$M^\Rm{eff}_{g \leftrightarrow e}$ between the ground state of the
magnetic configuration in the interacting region and  excited states
for the same number of particles. In particular, for the linear trimer, while
the ground state for three particles is an odd doublet, the most relevant
excited state for three particles is an even doublet. If the approximation
is valid, the effective matrix element should be smaller than the difference
between the energy of these states
\eqarray{
|E^{(3)}_{e}-E^{(3)}_g|& \gg &M^\Rm{eff}_{g \leftrightarrow e},
}{validez}
where
\begin{widetext}
\eqarray{
M^\Rm{eff}_{g\leftrightarrow e}&=&\sum_{j,\nu,n=2,4}\frac{1}{2}\left[\frac{\la \psi^{(3)}_g|H_\Rm{mix}| \psi^{(n)}_{j,\nu}\ra \la \psi^{(n)}_{j,\nu}|H_\Rm{mix}| \psi^{(3)}_{e}\ra}{E^{(n)}_{j,\nu}-E^{(3)}_g} + \frac{\la \psi^{(3)}_e|H_\Rm{mix}| \psi^{(n)}_{j,\nu}\ra \la \psi^{(n)}_{j,\nu}|H_\Rm{mix}| \psi^{(3)}_{g}\ra}{E^{(n)}_{j,\nu}-E^{(3)}_e}\right],\nonumber\\
}{meff}
\end{widetext}
is the effective matrix element
between the ground state $| \psi^{(3)}_{g}\ra$ and an excited state
$| \psi^{(3)}_{e}\ra$.
To estimate this matrix element in the case of the linear trimer, for simplicity, we have restricted the
sum above
to the lowest states with 2 and 4 particles in the EHSC.
The simplified expression is
\eqarray{
|E^{(3)}_{e}-E^{(3)}_g| &\gg& V^2 \alpha_{3g,2g} \alpha_{2g,3e}\times \nonumber\\
&&\times \left(\frac{1}{E^{(2)}_g - E^{(3)}_g}+ \frac{1}{E^{(2)}_g - E^{(3)}_e}\right),
}{validez2}
with $ \alpha_{a,b}$ being the generalization of Eq. (\ref{alphabeta})
for any pair of states $|a\ra$ and $|b\ra$. For the parameters used
in Fig. \ref{3d1d}, the largest value of
$M^\Rm{eff}_{g \leftrightarrow e}/|E^{(3)}_{e}-E^{(3)}_g|$ is $0.16$
for $\Delta/\pi U=4.0 \times 10^{-2}$.

In the case of the isosceles triangle the above criterion is taken
at the value of $V_g$ which renders $E^{(2)}_g=E^{(4)}_g$. For the
parameters in Fig. \ref{cond_isosc1}, the largest value of
$M^\Rm{eff}_{g \leftrightarrow e}/|E^{(3)}_{e}-E^{(3)}_g|$ is
$3.3\times 10^{-2}$ for $\Delta/\pi U=3.2 \times 10^{-2}$; and for
Fig. \ref{cond_isosc2}, the corresponding value is $0.29$ for
$\Delta/\pi U=1.6 \times 10^{-1}$.
Thus, we conclude that this set of
parameters is quite consistent with the approximation.

Quantitative details of our results are also affected by two other
approximations related with our slave-boson treatment: i) the neglect of
the triplet states in the subspaces of $N \pm 1$ particles in the interacting region,
and ii) the fact that the saddle point chooses the most convenient solution
between those with definite parity in these subspaces:
either  $e_{j+}=d_{j+}=0$ or $e_{j-}=d_{j-}=0$ for all $j$. However, in the
Kondo regime of most interest, the model reduces to a one-impurity
Kondo model, and for the linear trimer, both approximations slightly modify the Kondo temperature and
in opposite directions.\cite{trimer} Therefore our conclusions are not affected.
In particular, the system remains a Fermi liquid with ideal conductance
in the EHSC.
For other problems and larger $V$, the effects of the triplets can be important
near the singlet-triplet degeneracy.\cite{bulka}
 
As argued before,\cite{oguri} if the system is a Fermi liquid, its properties at
low-energy are expected to be the same as those of a non-interacting system, and
therefore, one expects non-zero conductance in the EHSC. Then, the possibility
of vanishing conductance in this case, seems to be related with the
breakdown of the Fermi liquid. For a trimer ($N=3$) with $C_{3v}$ symmetry,
a non-Fermi liquid ground state results as a consequence of the degeneracy
between odd and even doublets in the ground state of the
interacting region.\cite{laza,inge} In this case, although Eq. (\ref{landauer2})
suggests a vanishing conductance, our approach breaks down because
Eq. (\ref{validez}) is not satisfied. However, the geometry of the system
is incompatible with ``triangular" $C_{3v}$ symmetry, and even a very small
deviation of this symmetry restores Fermi liquid behavior.\cite{trimer,laza}
Therefore the most likely explanation of the result of B\"usser et al. \cite{busser}
of vanishing conductance in the EHSC for odd $N>1$ seems to be a failure
of the embedding procedure for $N>2$, in spite of its success when only one
QD is involved.\cite{torio,torio2,wily,chi}.
Recently, the case $N=3$ has been studied by NRG \cite{ogu3d}
and by Monte Carlo
and variational techniques.\cite{zit2} These results agree with ours
for small $V$.

\appendix

\section{The saddle-point approximation}\label{sec:spa}

This is a mean field approximation in which the problem reduces to
minimize the effective free energy $\tilde{F}$ as a function of
the {\em numbers} $e_{j,\nu}, d_{j,\nu}$ and $s_{\sigma}$. From Section \ref{sec:sbformalism}, $\tilde{F}$ is given by
\eq{
\tilde{F}=\tilde{F}_\Rm{boson} + F_\Rm{fermion},
}{Ft}
where
\begin{widetext}
\eqarray{
\tilde{F}_\Rm{boson}&=&\sum_{j,\nu} \left[(E^{(2)}_{j,\nu}+\lambda')\ e^2_{j,\nu} \right]+
\sum_{\sigma} \left[(E^{(3)}+\lambda'-\lambda_{\sigma})\ s^2_{\sigma} \right]+ \sum_{\sigma,j,\nu} \left[(E^{(4)}_{j,\nu}-2E^{(3)}+\lambda'-2 \lambda_{\sigma})\ d^2_{j,\nu} \right]-
\lambda',}{Sb}
\end{widetext}
and
\eqarray{
F_\Rm{fermion}&=&\sum_{\sigma} -\frac{1}{\beta}\int^{\infty}_{-\infty} \ln{\left(1+e^{-\beta (\omega-\mu)} \right)}\rho_{\sigma}(\omega)\ d\omega,\nonumber\\
}{Ff}
where $\rho_{\sigma}(\omega)$ is the total fermion density of states given by
\eqarray{
\rho_{\sigma}(\omega)&=&-\frac{1}{\pi}\Rm{Im\ Tr}\ \Bf{G}_{\sigma}(\omega),
}{rhof}
where the matrix $\Bf{G}_{\sigma}(\omega)$ contains the retarded Green functions
of the fermionic effective problem.

It is convenient to separate the fermionic free energy into a part
$F^0_\Rm{fermion}$
corresponding to the system without the effective site or ``impurity" (and therefore
independent of the bosonic variables) and the rest, which represents
the effect of the ``impurity" on $F_\Rm{fermion}$
\eqarray{
F_\Rm{fermion}&=&F^0_\Rm{fermion}+\Delta F_\Rm{fermion}.
}{Ff2}
The same separation can be made for the density of states. Using
the relation \cite{hew}
\eq{
\Delta \rho_{\sigma}=\frac{1}{\pi}\Rm{Im}\frac{\partial}{\partial \omega}\ln{G_{ff \sigma}}(\omega)
}{Drhof}
the change in free energy after adding the effective site becomes
\begin{widetext}
\eqarray{
\Delta F_\Rm{fermion}&=&-\sum_{\sigma}\frac{1}{\beta}\int^{\infty}_{-\infty} \ln{\left(1+e^{-\beta (\omega-\mu)} \right)} \frac{1}{\pi}\Rm{Im}\frac{\partial}{\partial \omega}\ln{G_{ff \sigma}}(\omega)\ d\omega,
}{DFf}
where $G_{ff \sigma}$ is the Green's function of the fermions $f$ at the effective
site.
Since the effective fermionic problem is non-interacting, $G_{ff \sigma}$
can be calculated easily using equations of motion.\cite{rev} The result is
\eqarray{
G_{ff \sigma}(\omega)&=&\lim_{\eta \rightarrow 0} \left[\omega + i\eta -\mu -E^{(3)}-\lambda_{\sigma} -  \sum_{\nu}\left( |V_{\nu,\sigma}|^2 g_{2,\nu,\sigma}(\omega)\right)\right]^{-1},
}{greenf}
\end{widetext}
where $g_{2,\nu,\sigma}(\omega)$ is the unperturbed Green's function (without
the effective site) at site 2 with parity $\nu$ and spin $\sigma$.

Integrating by parts in Eq. (\ref{DFf}) and taking $\mu=0$, one obtains
\eq{
 \left.\Delta F_\Rm{fermion}\right|_{T=0}=-\sum_{\sigma} \frac{1}{\pi}\Rm{Im}\int^0_{-\infty}\ln{G_{ff \sigma}}(\omega)\ d\omega\ ,}{DFf2}

Minimizing the effective free energy, one obtains the following equations
\begin{widetext}
\begin{eqnarray}
0&=&\frac{\partial \tilde{F}}{\partial e_{j,\nu}},\nonumber\\
0&=&e_{j,\nu}\left(\lambda' +E^{(2)}_{j,\nu} \right)-\sum_\sigma \frac{\sqrt{2}V s_\sigma}{\pi}\Rm{Im} \left[V_{\nu,\sigma}\alpha_{j,\nu}\int^0_{-\infty} d\omega \ g_{2,\nu,\sigma}(\omega)G_{ff \sigma}(\omega)\right],\label{e}\\
0&=&\frac{\partial \tilde{F}}{\partial d_{j,\nu}},\nonumber\\
0&=&d_{j,\nu}\left(\lambda' +E^{(2)}_{j,\nu}-2E^{(3)}-\sum_\sigma\lambda_\sigma \right)- \sum_\sigma \frac{\sqrt{2} V s_\sigma}{\pi}\Rm{Im} \left[ V_{\nu,\sigma} \alpha_{j,\nu}\int^0_{-\infty} d\omega \ g_{2,\nu,\sigma}(\omega) G_{ff \sigma}(\omega) \right],\label{d}\\
0&=&\frac{\partial \tilde{F}}{\partial s_\sigma},\nonumber\\
0&=&s_\sigma\left(\lambda' -\lambda_\sigma \right)-\frac{\sqrt{2}V }{\pi}\Rm{Im} \left\{ \sum_{j,\nu} \left[ V_{\nu,\sigma} \left( \alpha_{j,\nu} e_{j,\nu} + \beta_{j,\nu} d_{j,\nu} \right)\int^0_{-\infty} d\omega \ g_{2,\nu,\sigma}(\omega) G_{ff \sigma}(\omega) \right]\right\},\label{s}\\
0&=&\frac{\partial \tilde{F}}{\partial \lambda'},\nonumber\\
0&=&\sum_{j,\nu}e^2_{j,\nu} + 2 s^2 + \sum_{j,\nu} d^2_{j,\nu} - 1,\label{lambda'}\\
0&=&\frac{\partial \tilde{F}}{\partial \lambda_\sigma},\nonumber\\
0&=&-\frac{\Rm{Im}}{\pi}\int^0_{-\infty} d\omega \ G_{ff \sigma}(\omega)- s^2_\sigma - \sum_{j,\nu} d^2_{j,\nu}.\label{lambda}
\end{eqnarray}
\end{widetext}
Since one does not expect breaking of SU(2) symmetry we take
$s_\uparrow = s_\downarrow = s$ and
$\lambda_\uparrow = \lambda_\downarrow = \lambda$.
For small $V$ we can take a constant unperturbed density of states
with its value at the Fermi energy $\rho_0=1/(2 \pi t)$.\cite{side}
This simplifies the calculations and does not
affect our conclusions.
\begin{equation}
g_{2,\nu}(\omega)=-i \pi \rho_0 \theta(D-|\omega|),
\label{gc}
\end{equation}
where $2 D \rho_0=1$ ($D= \pi t)$.

For $N=3$, there are 15 independent variables. However, with a little algebra
the number of independent variables can be reduced to four.
Using Eq. (\ref{e}), we can relate all boson numbers $e_{j,\nu}$ in terms
of one of them. We choose $e_{0,\nu}$ corresponding to the ground state
of $N-1$ particles:
\eqarray{
e_{j,\nu}&=&\frac{\alpha_{j,\nu}}{\alpha_{0,\nu}}\left(\frac{E^{(2)}_{0,\nu}+\lambda'}{E^{(2)}_{j,\nu}+\lambda'}\right) e_{0,\nu},\nonumber\\
&=&R^{(2)}_{j,\nu} \ e_{0,\nu},
}{ej}
where we have defined
\eqarray{
R^{(2)}_{j,\nu}&\equiv&\frac{\alpha_{j,\nu}}{\alpha_{0,\nu}}\left( \frac{E^{(2)}_{0,\nu}+\lambda'}{E^{(2)}_{j,\nu}+\lambda'}\right).
}{r2}

Similarly, using Eqs. (\ref{e}) y (\ref{d}) all  $d_{j,\nu}$ can also be
related to $e_{0,\nu}$ giving
\eqarray{
d_{j,\nu}&=&\frac{\alpha_{j,\nu}}{\alpha_{0,\nu}}\left( \frac{E^{(2)}_{0,\nu}+\lambda'}{E^{(4)}_{j,\nu}+\lambda'-2E^{(3)}-2\lambda}\right) e_{0,\nu},\nonumber\\
&=&R^{(4)}_{j,\nu} \ e_{0,\nu},
}{dj}
where
\eqarray{
R^{(4)}_{j,\nu}&\equiv&\frac{\alpha_{j,\nu}}{\alpha_{0,\nu}}\left( \frac{E^{(2)}_{0,\nu}+\lambda'}{E^{(4)}_{j,\nu}+\lambda'-2E^{(3)}-2\lambda}\right).
}{r4}

Using these relations in Eq. (\ref{e}), we can write
\begin{widetext}
\eq{
0=e_{0,\nu}\left\{ \left(\lambda' +E^{(2)}_{0,\nu} \right) -  2 V^2 s^2 \rho_0 \alpha_{0,\nu} \left[\sum_j \left( R^{(2)}_{j,\nu}+ R^{(4)}_{j,\nu}\right) \alpha_{j,\nu}\right] \ln{\left[\frac{\left(E^{(3)}+\lambda\right)^2+\delta^2 }{\left(E^{(3)}+\lambda+W\right)^2+\delta^2}\right]}\right\},
}{e2}
where $\delta=\pi \rho_0 \left(|V_-|^2+|V_+|^2\right)$.
Eq. (\ref{lambda}) takes the form
\eq{
0=\frac{1}{2}\left[\sum_{j,\nu} e^2_{0,\nu} \left( \left(R^{(2)}_{j,\nu}\right)^2- \left(R^{(4)}_{j,\nu}\right)^2 \right) \right]-\frac{1}{\pi}\tan^{-1}\left[\frac{E^{(3)}+\lambda}{\delta}\right].
}{lambda2}

Finally (\ref{s}) becomes
\eqarray{
0&=&1+ \frac{V^2 \rho_0 \left(\sum_\nu \left[\sum_j e_{0,\nu} \alpha^{(0)}_{j,\nu} \left( R^{(2)}_{j,\nu}+ R^{(4)}_{j,\nu}\right)\right]^2\right)}{\lambda'-\lambda}  \ln{\left(\frac{\left(E^{(3)}+\lambda\right)^2+\delta^2 }{\left(E^{(3)}+\lambda+W\right)^2+\delta^2}\right)}.
}{s2}
\end{widetext}
In practice, Eq. (\ref{lambda'}) has been used to express  $s$ in terms
of the other bosonic variables and Lagrange multipliers. This leads to a system
of four coupled non-linear equations with the unknowns
$\{ e_{0,+}, e_{0,-},\lambda,\lambda'\}$, which was solved numerically.
In addition, one of the $e_{0,\nu}$ should vanish to satisfy both
Eqs. (\ref{e2}). For the trimer  $e_{0,-}=0$.
To facilitate the numerical solution, we started solving the equations
in the EHSC, for which  $e_{j,\nu}=d_{j,\nu}$, and then
$\lambda=E^{(3)}$, as can be easily seen from Eqs. (\ref{e}) y (\ref{d}).
After the solution in this case was obtained, the resulting values were used
as a starting guess for a slightly modified $V_g$ and the process was repeated
until the whole range of gate voltages of interest was covered.

\section{Analytical results for one dot}

\label{sec:verification_fsr}

If the interacting system is composed of only one dot, in the absence of a
magnetic field $\mathbf{B}$, the change in the effective free energy in the
saddle-point approximation depends on two independent variables and the
problem is simplified considerably.

For $\mathbf{B}=0$, $s_{\uparrow }=s_{\downarrow }=s$. Eliminating $e$ from
the constraint $e^{2}+2s^{2}+d^{2}=1$ [see Eq. (\ref{constraint1})], the
change in the effective free energy can be written as
\begin{eqnarray}
\Delta \tilde{F}^{1D} &=&\tilde{F}_{\Rm{boson}}^{1D}+\Delta F_{\Rm{fermion}%
}^{1D},  \nonumber \\
\tilde{F}_{\Rm{boson}}^{1D} &=&-2\lambda \ s^{2}+\left( U-2\lambda \right) \
d^{2},  \nonumber \\
\Delta F_{\Rm{fermion}}^{1D} &=&-\frac{2}{\pi }\mathrm{Im}\int_{-\infty
}^{0}\ln G_{ff \sigma }^{1D}(\omega )\ d\omega \ ,  \label{f1d}
\end{eqnarray}
with
\begin{equation}
G_{ff\sigma }^{1D}(\omega )=\left[ \omega -(E_{d}+\lambda )+i\delta \right]
^{-1},  \label{g1d}
\end{equation}
where the half width of the resonance is

\begin{equation}
\delta =\Delta s^{2}(d+e)^{2}.  \label{del}
\end{equation}

As in Ref. \onlinecite{paula},  $\lambda $ can be expressed in terms of $s$ and $d$
using
\begin{equation}
s^{2}+d^{2}=\langle n_{f\sigma }\rangle =-\frac{1}{\pi }\mathrm{Im}%
\int_{-\infty }^{0}G_{ff\sigma }^{1D}=\frac{1}{\pi }\tan ^{-1}{\left( \frac{%
\delta }{E_{d}+\delta }\right) },  \label{dDF1d}
\end{equation}
giving
\begin{equation}
\lambda (s,d)=\frac{\delta }{\tan {\left[ \pi (s^{2}+d^{2})\right] }}-E_{d}.
\label{elimlambda}
\end{equation}

The two-terminal Landauer formula at $T=0$ gives
\begin{equation}
G=\frac{2e^{2}}{h}\delta ^{2}|G_{ff \sigma}^{1D}|_{\omega =0}^{2}=\frac{2e^{2}}{h}%
\sin ^{2}{\left( \pi \langle n_{f\sigma }\rangle \right) },
\label{landauer1d}
\end{equation}
where in the second equality, Eqs. (\ref{g1d}) and (\ref{elimlambda}) were
used. Thus, Eq. (\ref{fsrc}) is verified.

Replacing Eqs. (\ref{g1d}) and (\ref{elimlambda}) in (\ref{f1d}) the
change in free energy becomes
\begin{widetext}
\begin{eqnarray}
\Delta \tilde{F}^{1D} &=&Ud^{2}+2E_{d}(s^{2}+d^{2})+\frac{1}{\pi }\left\{ \ln \left[ \frac{\gamma ^{2}+\delta ^{2}}{(D+\gamma
)^{2}+\delta ^{2}}\right] -2(D+\gamma )\arctan \left( \frac{\delta }{%
D+\gamma }\right) \right\},  \label{fd}
\end{eqnarray}
\end{widetext}
where $\gamma =\delta \cot [\pi (s^{2}+d^{2})]$. In the EHSC, $%
e^{2}=d^{2}=1/2-s^{2}$, and the energy depends on only one variable $s$,
which can be obtained   by minimization. In the Kondo limit of large $U$,
$d^{2}\rightarrow 0$, the resulting transcendental equation can be simplified
giving

\[
d^{2}\simeq \frac{D}{4\Delta }\exp \left( \frac{-\pi U}{4\Delta }\right) ,
\]
and using Eq. (\ref{del})

\begin{equation}
T_{K}\simeq \delta \simeq \frac{D}{2}\exp \left( \frac{-\pi U}{4\Delta }%
\right) ,  \label{tks}
\end{equation}
which has the correct exponential dependence. If instead, the mean-field
approach of Ref. \onlinecite{kr} is used (including some suitable roots to
reproduce the non-interacting case), an additional factor 4 appears in the
denominator of the exponent in Eq. (\ref{tks}).

\acknowledgments

We are partially supported by CONICET. This work was sponsored by
PICT 03-13829 of ANPCyT.
\bibliographystyle{apsrev}

\end{document}